\documentclass[useAMS,aps,prl,onecolumn,showpacs,superscriptaddress,groupedaddress]{revtex4}  
\usepackage{graphicx}  
\usepackage{dcolumn}   
\usepackage{bm}        
\usepackage{amssymb}   
\usepackage[normalem]{ulem}
\usepackage[T1]{fontenc}
\usepackage[usenames]{color}

\def\kms{{\,\rm km\,s^{-1}}}
\def\be{\begin{equation}}
\def\ee{\end{equation}}

\newcommand{\beq}{\begin{equation}}
\newcommand{\beqa}{\begin{eqnarray}}
\newcommand{\bey}{\begin{eqnarray}}
\newcommand{\eeq}{\end{equation}}
\newcommand{\eey}{\end{eqnarray}}
\newcommand{\eeqa}{\end{eqnarray}}

\def\lsim{\mathrel{\raise.3ex\hbox{$<$\kern-.75em\lower1ex\hbox{$\sim$}}}}
\def\gsim{\mathrel{\raise.3ex\hbox{$  $\kern-.75em\lower1ex\hbox{$\sim$}}}}

\def\kms{\, {\rm km \, s}^{-1} }

\def\rho{{n}}

\begin{document}
\title{Gravity theories, Transverse Doppler and Gravitational Redshifts in Galaxy Clusters}
\author{HongSheng Zhao}
\affiliation{Scottish University Physics Alliance, University of St Andrews, KY16 9SS, UK}
\author{John A. Peacock}
\affiliation{Scottish University Physics Alliance, University of Edinburgh, EH9 3HJ, UK}
\author{Baojiu Li}
\affiliation{Institute of Computational Cosmology, Department of Physics, Durham University, Durham DH1 3LE, UK}
\date{1.1.2012; Email address: hz4@st-andrew.ac.uk}
\begin{abstract}
There is growing interest in testing alternative gravity theories
using the subtle gravitational redshifts in clusters of galaxies.
However, current models all neglect a transverse Doppler redshift of
similar magnitude, and some models are not self-consistent.  An
equilibrium model would fix the gravitational and transverse Doppler
velocity shifts to be about $6\sigma^2/c$ and $3\sigma^2/2c$ in
order to fit the observed velocity dispersion $\sigma$
self-consistently.  This result comes from the Virial Theorem for a
spherical isotropic cluster, and is insensitive to the theory of
gravity.  A gravitational redshift signal also does not directly
distinguish between the Einsteinian and $f(R)$ gravity theories,
because each theory requires different dark halo mass function to keep the clusters  
in equilibrium. When this constraint is imposed, 
the gravitational redshift has no sensitivity to theory.  {\bf Indeed our N-body simulations show that 
the halo mass function differs in $f(R)$, and that 
the transverse Doppler effect is stronger than analytically predicted due to non-equilibrium.}
\end{abstract}
\maketitle


The theory of gravity has been subjected to various tests with the
ever-improving quality of data from cosmology, galaxy clusters,
galaxies and the solar system \cite{manypapers}.  As shown by recent
numerical N-body simulations on $f(R)$ type or scale-coupled
gravities \cite{nbodypapers}, dynamical data on non-linear cluster
scales help to break theoretical degeneracies on linear cosmological
scales, and overcome statistical uncertainties in observations.  Past
techniques often proposed comparing lensing data and kinematic data with
simulations \cite{ZhaoLiK}, which can involve significant amounts of effort in
modeling of the mass distribution before indirect constraints can be set
on the gravitational potential $\Phi(r)$ of the cluster.  It
would clearly be better to measure the gravitational potential
in a galaxy cluster directly and compare this potential with the
prediction from the Poisson equation for the mass distribution in a
given gravity theory.

Indeed the gravitational potential is an observable from the shift of 
spectral lines \cite{Broadhurst}.  Lines from the surface of the Sun, e.g., are shifted by 
$GM_\odot/R_\odot c \simeq 0.6\kms$, and more for compact stars.  On cosmic
scales, the deepest potential well $\Phi(r)$ is felt by the bright central galaxy (BCG) in a cluster of 
galaxies, where a nearly spherical distribution of many hundreds of
galaxies orbit around the centre, with a Gaussian dispersion of random
velocities of $\sigma(r) \sim 1000\kms$  in each direction.  The
observed line-of-sight Doppler shifts of galaxies relative to the BCG
satisfy a Gaussian distribution with a small but non-zero mean
velocity.  This is partly due to the gravitational redshift (GR), a
feature in any metric theory of gravity, caused by the spatial variation of the
gravitational potential:
\be
\Delta_{\rm GR}=\left[\Phi_{\rm BCG}-\Phi(r)\right]/c. 
\ee 
This signal of $\sim 10\kms$ becomes detectable above the
$\sigma/\sqrt{N}$ uncertainty of the mean velocity once the sample
size $N \ge 10^4$.  To obtain such a large sample for the first time,
Wojtak et al.~ \cite{W11} used $N \sim$ 125,000 galaxies from about 7,800 
clusters from the Sloan Digital Sky Survey (SDSS), divided the
galaxies into four bins according to their projected distances $R$
from their respective BCGs, "stack" their light-of-sight velocities
relative to their BCGs, carefully removed interlopers, and finally
computed the mean velocity in each bin.  In this paper, we
investigate the pros and cons of the gravitational redshift approach,
and for the first time introduce a new effect in galaxy clusters.

In fact, the gravitational redshift is supplemented by an additional
redshift of comparable amplitude.  For any metric theory of gravity
 \cite{manypapers} the space time near a galaxy cluster is described by
the metric $d\tau^2 = (1+2\Phi/c^2) dt^2 - (1+z)^{-2} (1+2\Psi/c^2)
dx^2$.  Light emitted from a cluster at redshift $z$ is time-dilated 
with the ratio of the observed wavelength to the emitted wavelength satisfying:
\begin{equation}
{c \over  (1+z)} {\lambda_{\rm obs} \over \lambda_{\rm emit}} = \left[c + {\Phi - {\mathbf v}^2/2 \over c}\right], 
\end{equation} 
{\bf which reveals an additional effective radial velocity shift}
\be 
\Delta_{\rm TD} = {\left[\left<|{\mathbf v}|^2\right> - |{\mathbf v}_{\rm BCG}|^2\right] / 2c}, 
\ee 
owing to the transverse Doppler (TD) effect from random motions of
galaxies in special relativity (SR).  

Wojtak et al.~ \cite{W11} reported a blueshifting
of the mean apparent line-of-sight velocity of the galaxies in the
SDSS clusters, again relative to the BCG, which was then interpreted
as purely GR.  But this interpretation is incomplete.  The TD effect always
co-exists in proportion to GR because of the Virial Theorem:
\begin{equation}
\left<{-\Phi / 2} \right>/2c = {\left<G M / r \right> /2c} = {\left<|{\mathbf v}|^2\right> /2c},
\end{equation} 
where {\bf $M$ is the mass enclosed within a radius $r$}, $\left<\right>$ denotes the averaging over all gravitational
masses in the whole virialized volume of a cluster, and the factor of
$1/2$ in front of $\Phi$ prevents double counting of the pairwise
mutual potential.  Thus the random kinetic energy per unit mass
$\overline{{\mathbf v}^2/2}$ is globally $25\%$ of the average
potential $\overline{-\Phi}$.  The ratio of $1/4$ holds even after
averaging over a distribution of clusters of different mass and for
clusters of any density profile and anisotropy parameter, so the
Virial Theorem is a robust link between TD and GR effects, and their
superposition is observed as the mean velocity shift.  
{\bf In Fig~1. we show the TD and GR shifts in haloes from the N-body simulations of \cite{ZhaoLiK}.  
As we can see that haloes tend to be blue-shifted at large radii compared to their centres due to combination of GR and TD effects. }
{\bf In reality, however, Fig.~1 shows that the TD effects are often 
enhanced by a factor 1 to 4 in halos in N-body simulations because their viral ratio 
often deviates from the expected value at virial equilibrium. }

\begin{figure}
\leftline{\includegraphics[angle=0,width=9cm]{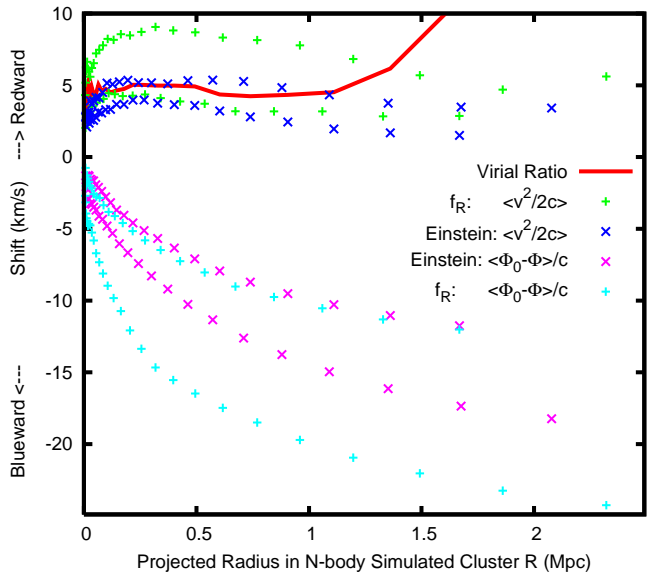}}
\caption{ 
Shows N-body simulations predicted TD and GR velocity shifts at projected radius R, {\bf annotated by $\left<v^2/2c\right>$ and $\left<\Phi(0)-\Phi(r)\right>/c$
}. {\bf Pluses} show the 20th and the 1st biggest haloes of $r_{\rm vir}=1.67$ and $2.32$ Mpc in $f(R)$  with $|f_{R0}|=10^{-4}$.
{\bf Crosses} show the 10th and the 1st  biggest haloes of $r_{\rm rvir}=1.67$ and $2.08$ Mpc in Einsteinian gravity. 
{\bf The red thick line} shows the {\it dimensionless} virial ratio $\left<v^2\right>/\left<Z\partial_Z\Phi\right>$ for typical halos in GR, which deviates from its equilibrium value 3, especially at larger radii, i.e., the TD effect is 1 to 4 times its equilibrium prediction. }\label{fig1}.
\end{figure}

It seems straightforward to test many gravity theories with their gravitational redshift
prediction.  However, {\bf recent tests of modified gravity often assume that clusters have dark halos}, which complicates the tests.  
E.g., {\bf Hu \& Sawicki \citep{hs2007} show in $f(R)$ gravity with $|f_{R0}|=10^{-4}$ 
the Newton constant $G$ is boosted by a nearly constant factor $4/3\simeq 1.33$ for typical halos 
on cluster scales}.  For a fixed cluster mass Wojtak et al. claimed that this 33\% boost of the gravitational
redshift signal robustly distinguishes Modified Gravity from
Einsteinian Gravity.  Such a claim, however, has a flaw: the total
mass of the dark halo is an unknown free parameter, which must be
determined by fitting the observed velocity dispersion as a function of
distance from the cluster centre.  Since $G$ appears only in the
combination $GM$, one can cancel essentially the enhancement of $G$ in
$f(R)$ gravity by reducing the halo mass parameter $M$, thus
obtaining the {\it indistinguishable} fit to the velocity dispersion curve and to the
mean velocity shift signal.  {\bf Nevertheless, 
one could test whether statistics of redshift data and halo counts is 
indeed biased towards more massive haloes, as generically found in 
$f(R)$ theory. Fig.~1 shows very massive haloes are more frequent in $f(R)$ gravity than in Einsteinian gravity, 
but for haloes of similar virial radius or virial velocity or $GM_{\rm vir}$, two theories predict a similar shift.}

A more specific example is to use the isotropic Jeans equation
$-GM/r^2= {d (\rho \sigma^2) / \rho dr}, $
where the tracers, i.e., galaxies are assumed an isotropic dispersion $\sigma(r)$ and a number density $\rho
\propto r^{-\gamma}$ at large radius $r$.  One solves for the random kinetic energy 
$\overline{{\mathbf v}^2}/2 = {1.5\sigma^2} = (1.5G M/r)/(\gamma + 1)$,
which is locally $\simeq 3/8-3/10$ of a Keplerian potential
$GM/r$ for a galaxy count profile with $\gamma \simeq
3-4$ at large radii.  The ratio $3/8$ or $3/10$ holds even after
stacking of clusters of different masses and line-of-sight
projection. This argument is true in standard gravity as well as in
$f(R)$ gravity.  Clearly gravity models with the same $GM$ predict the
same dispersion curve, and velocity shifts.

To compute the TD and GR effects generally at any projected radius, we
start with the isotropic Jeans equation
$-{\partial (\rho \sigma^2) / \partial Z} =  \rho {\partial \Phi / \partial Z} $
for the observable tracers (galaxies) with a number density $\rho(r)$
in equilibrium in the potential $\Phi(r)$.  We integrate this 
over the line-of-sight depth $Z$ through a cluster after multiplying
by $Z dZ$, and apply 
an integration by parts to $ Z d (\rho \sigma^2)$
to drop the total derivative term.   We find
$\int_{-\infty}^{\infty} dZ \left(\rho \sigma^2\right) = \int_{-\infty}^{\infty} dZ  \left(\rho Z {\partial \Phi / \partial Z}\right),$
which predicts that inside the virial radius the specific 3D kinetic energy averaged in a projected annulus $R$ to $R+dR$ is 
${\left<|{\mathbf  v}|^2\right> / 2} = \left< {GM(r) Q Z^2/r^{3}} \right>$, 
where $Q=3/2$ from quadrature sum of the three velocity components.
This expression allows us to predict the SR effect at all radii for any
matter density in any metric-based gravity theory, since the Jeans
equation applies to any force which is a gradient of a potential.
E.g., if the density is $\sim r^{-\gamma} \sim r^{-3} $ and gravity $\sim
r^{-2}$ then
${\left<v^2\right> / 2}  = (3 / 8)\left<|\Phi|\right>$, where  $\left<|\Phi|\right>=(\pi GM / 4R)$ 
is the density-weighted line-of-sight integration of $|\Phi|$.

\begin{figure}
\leftline{\includegraphics[angle=0,width=9cm]{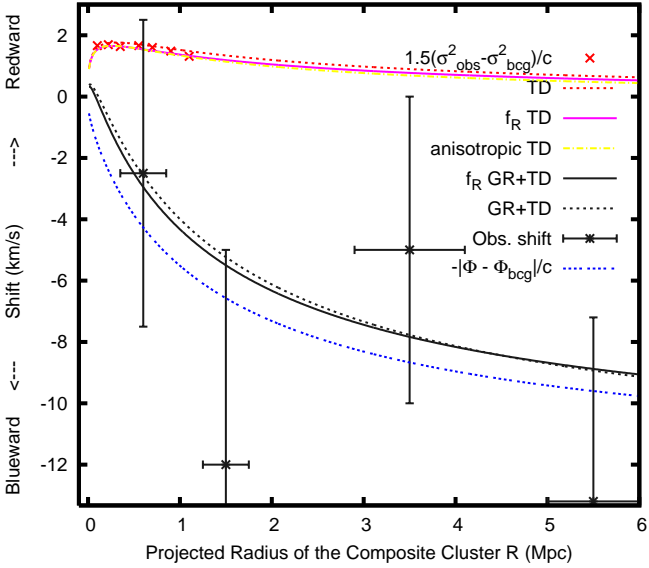}}
\caption{ 
Shows consistency of the data (crosses and error bars) \cite{W11} with the competing
GR effect $-[\smash{\overline{\Phi}} -\Phi_{\rm BCG}]/c$ plus TD effect $ {(\smash{\overline{{\mathbf v}^2}} - {\mathbf v}_{\rm BCG}^2) / 2c}$;   
the 3D kinetic energy 
 ${\smash{\overline{{\mathbf v}^2}} / 2}$ at any $R$ equals $\smash{\overline{Q Z\partial \Phi/\partial Z}}$ averaged over the line of sight depth $z$, where $Q=3/2$ within the virial radius $r_{\rm vir}$ of an isotropic cluster, except  
$Q \simeq (3r_{\rm   vir} +r)/(2r_{\rm vir} +r)$ for the yellow line to mimic mild anisotropy and non-equilibrium at large radii.  
Clusters are all modelled as NFW haloes, weighted by $M_{\rm vir}^{-7/3}$
in the virial mass range $M_{\rm vir}=(0.11-2) \times 10^{15}M_\odot$ in Einsteinian gravity (dashed), or 
$M_{\rm vir}=(0.09-1) \times 10^{15}M_\odot$ in $|f_{R0}|=10^{-4}$ gravity (solid) with a 33\% boost of the effective G 
in these lightish haloes (Fig. 3 of \cite{Sch10}).
}\label{fig2}.
\end{figure}

To compute the GR and TD effects for the SDSS clusters, we account for different cluster   
masses using a Salpeter-like mass function $dN/dM_{\rm vir} \sim
M_{\rm vir}^{-2.33} \sim M_{\rm vir}^{-7/3}$ between the mass
range $M_l$ and $M_u$ as Wojtak et al., so 
\begin{equation}
\left<|{\mathbf v}|^2/2\right> = \overline{(Q Z^2/r)(d \Phi/d r)}/\overline{1}, ~ \left<{\Phi_0-\Phi}\right>=\overline{(\Phi_0-\Phi(r))} /\overline{1}, 
~~\overline{A}\equiv \int_{M_l}^{M_u} dM_{\rm vir}  \int_{-\infty}^{\infty}  dZ  \left. \rho(r) A\right|_{r=\sqrt{Z^2+R^2}},
\end{equation}
%
where {\bf $\overline{A}$ is essentially a stacked density-weighted line-of-sight integration of a quantity $A$ at the projected radius $R$}, and 
the spherical potential and tracer (galaxy) number count density are given by 
$\Phi(r) = - {G  M_{\rm vir} \over r F(C)}  \ln ( 1 + {r C / r_{\rm vir}} ),$ and $\rho(r) \propto {M_{\rm vir}^{-7/3} \over 4 \pi F(C)} {N_{\rm vir} \over r ( r +   r_{\rm vir}/C)^2}$,
where $F(x) \equiv \left[ \ln(1+x) - x/(1+x) \right]$.  We fix
the halo concentration parameter $C=5$ and the virial radius $r_{\rm vir} 
= 1.2 \left(M_{\rm vir}/10^{14}M_\odot\right)^{1/3}$, as determined in \cite{W11}.  
Such a spherical NFW potential is an approximation to the
true potential in Einsteinian gravity; in $f(R)$ gravity the potential
from N-body simulations tends to be more concentrated \cite{ZhaoLiK}
and in TeVeS gravity the potential tends to have a pure $\ln(r)$
profile at large radii.  A reasonable analytical approximation of 
the tracer (galaxy) count $n(r)$ is the spherical NFW profile with the count of galaxies inside the
virial radius $N_{\rm vir} \propto M_{\rm vir}$.  
Here we do not attempt to model the detailed selection criteria of galaxies of measurable redshift
and the off-centredness of the BCGs since these complexities seem not to be the fundamental issue here.  
We do model the effects of mild anisotropy and non-equilibrium at large radii. 
From N-body simulations of \cite{ZhaoLiK,Maccio} we find that $Q \simeq
(3r_{\rm vir} +r)/(2r_{\rm vir} +r) \le {3/2}$ works well empirically. 

The results for Einsteinian gravity gravity are shown in Fig.~2 for 
a halo mass range of $(M_l,M_u)=(0.11\times 10^{15}, 2\times
10^{15}) M_\odot$.  Note these fitting parameters are deduced
from hydrostatic balancing of the pressure gradient  
${d(\rho \sigma^2_{\rm obs}) /(\rho dr)}$ and the halo gravity, as one cannot
directly observe the halo and measure its mass.  One can see our
choice of parameters can fit the observed ${(3 \sigma_{\rm obs}^2 - 3
  \sigma_{\rm BCG}^2) / 2c}$ curve of \cite{W11} and, in doing so, we can predict
the $-|\Phi(R)-\Phi_{\rm BCG}^2)|/c$ GR curve.  Note the inevitable
{\it reversal} from the observed average $\simeq 6.5 \pm 4 \kms$
blueshifting to redshifting when within 0.2 Mpc of the BCG (cf. grey
error bars and lines in Fig.~2) due to TD: the line-of-sight
dispersion of non-BCGs $\sigma_{\rm obs}(R) \simeq 600\kms \simeq
3\sigma_{\rm BCG} $ converts directly to a ${(3 \sigma_{\rm obs}^2 - 3
  \sigma_{\rm BCG}^2) / 2c} \simeq 1.6 \kms$ TD differential redshift
near an isotropic cluster centre.  The TD signal (red crosses) is
clearly both non-negligible and model-insensitive, and is thus a robust
constraint applicable to any metric-based gravity theory.

As stated earlier, one should not compare an $f(R)$ gravity model with 
an Einsteinian gravity model of the same halo mass distribution,
namely $(M_l,M_u)=(0.11\times 10^{15}, 2\times 10^{15}) M_\odot$,
since it would overpredict the velocity dispersion curve $\sigma^2(R)$
everywhere by the same factor of $4/3$, which can be ruled out without
even measuring gravitational redshift.  In fact the isotropic
Jeans equation ensures a {\bf one-to-one} relation between the SR and GR effects.  
The mass distribution where all (halo virial) mass is lowered by the same factor $4/3$ would predict a
velocity dispersion curve identical as the Einsteinian curve.
Instead, to show some difference, here we adopt $(M_l,M_u)=(0.09\times
10^{15}, 1\times 10^{15}) M_\odot$, and the result is shown as solid
lines in Fig.~2.  This $f(R)$ model produces GR and
TD shifts by amounts essentially identical to Einsteinian gravity.
Thus GR, TD and the velocity dispersion profile contain essentially
three redundant copies of information about a metric theory, up to
some uncertainty from anisotropy.

Likewise, the claimed $(0-10)\kms$ extra shift in TeVeS reduces to
only $(0-3)\kms$ when adopting mass models consistent with
$\sigma_{\rm obs}^2$ \cite{BekS}.  Unfortunately the TD effect is
left out explicitly in {\it all\/} these papers: e.g., TeVeS predicts a
roughly radius-independent SR red-ward shift of $\left<{ 
(3Z/2c) \partial \Phi(r) / \partial Z} \right> = 3\sigma_\infty^2
/ \gamma c \simeq 1\kms$ for $\gamma \sim 3$.  Further
investigation including all relativistic effects in $f(R)$ N-body halo simulations
\cite{ZhaoLiK,Sch10} would be needed.

{\bf While we cannot break degeneracy of gravity theory as long as different effective $G$ and halo mass $M$ yield the same virial velocity}, 
the differential shift in a cluster is a {\bf remarkably sensitive}  
measure of the mass function of haloes {within the standard gravity}.  It can be easily shown that,
the global gravitational shift with respect to the center, integrated inside {\bf an aperture of $R \rightarrow \infty$}  
and averaged over all haloes, is given by 
\be
\left<c \Delta_{GR} \right> \equiv  {\int V_{vir}^2 dN \over \int dN}  \left<f\right> 
=  {(3\alpha-3) [ 1 - (V_{vir,l}/V_{vir,u})^{3\alpha-5}]  V_{vir,u}^2 \over (3\alpha - 5) [ 1 - (V_{vir,l}/V_{vir,u})^{3\alpha-3}]~~~~~}{C \over F(C)},  
\ee
{\bf where 
$f(r) \equiv {\Phi(0) - \Phi(r)\over V_{vir}^2}= [-1 + {r_s \over r} \ln(1+{r \over r_s})] {C \over F(C)}$ is a rescaled NFW potential,
$\left<f\right> = {C \over F(C)} \sim 5$ is its density weighted global average for halos of typical concentration $C \sim 5$.  Here the mass function $dN \sim M_{vir}^{-\alpha} dM_{vir} \sim V_{vir}^{-3\alpha+2}dV_{vir}$ for the virial velocity between the lower bound $V_{vir,l}$ and upper bound $V_{vir,u}$. For a fixed power-law index $\alpha \sim 7/3$, the differential shift $\left<c \Delta_{GR} \right> \sim 10 V_{vir,u}^2$, which is an indirect measurement of the virial velocity $V_{vir,u}$ at upper mass cutoff. }

Note finally the future possibility of other relativistic effects\cite{Broadhurst}, e.g., measuring 
the GR and TD effects from cluster X-ray gas spectra.  This has
less need for stacking clusters because of negligible $\sigma/\smash{\sqrt{N}}$ uncertainty 
for the countless ionized gas particles.  The signal also differs from the
technique of\cite{W11} because the X-ray gas particles have a
profile different from that of galaxy number density, and less
velocity anisotropy than galaxies.  By comparing
the transverse Doppler signals of different tracers one could even infer the velocity anisotropy of galaxies inside clusters.

{}

\vfill
\eject

\end{document}